\documentclass[prb,twocolumn,showpacs]{revtex4}

\usepackage{graphicx}
\usepackage{dcolumn}

\newcommand{\beq}{\begin{equation}}
\newcommand{\eeq}{\end{equation}}
\newcommand{\beqa}{\begin{eqnarray}}
\newcommand{\eeqa}{\end{eqnarray}}

\begin{document}

\title{Band insulator to Mott insulator transition in a bilayer Hubbard model}
\author{S. S. Kancharla and S. Okamoto} \affiliation{Materials Science
and Technology Division, Oak Ridge National Laboratory, Oak Ridge, TN
37831}
\begin{abstract}
The ground state phase diagram of the half-filled repulsive Hubbard
model in a bilayer is investigated using cluster dynamical mean field
theory. For weak to intermediate values of Coulomb repulsion $U$, the
system undergoes a transition from a Mott insulating phase to a
metallic phase 
and then onto a band insulating phase as the interlayer hopping is
increased. In the strong coupling case, the model exhibits a direct
crossover from a Mott insulating phase to a band insulating
phase. These results are robust with respect to the presence or
absence of magnetic order.
\end{abstract}
\pacs{71.10.-w, 71.10.Fd, 71.10.Hf, 71.27.+a, 71.30.+h, 71.45.Lr} 
\date{\today}
\maketitle

Strong Coulomb interaction can localize electrons even in the absence
of disorder to give rise to a gapped electronic state known as the
Mott insulator. This state is of tremendous interest because upon
addition of carriers it produces high temperature superconductivity
and intriguing features such as a pseudogap in the normal state of the
cuprates. Band insulators, on the other hand, lack the ability to
conduct because they have either a filled or an empty band. The topic
of transitions between band and Mott insulators has received
considerable attention
recently\cite{sarma.prl,scalettar.prl,fuhrmann,aligia,kampf,noack,garg},
because it provides a useful tool to understand the fundamental
differences between the two states. Recent progress in the control of
Fermionic atoms trapped in optical lattices adds to the experimental
interest as it would be feasible to investigate the evolution of
correlated ground states in a many body system. In a recent study, it
was shown that the ground state of a Mott-insulating Haldane phase in
a generalized Hubbard ladder can be connected {\it adiabatically} to a
band insulating phase\cite{rosch}. On the other hand, studies of the two
dimensional ionic Hubbard model (IHM), which contains an alternating
local potential $\pm \Delta$ on the A or B sublattice, find an
intervening phase between the band and Mott insulating phases rather
than a continuous evolution when the Coulomb repulsion is
varied. Cluster dynamical mean field theory (CDMFT) calculations for
the IHM at zero temperature suggest that the intervening phase is a
bond-ordered insulator\cite{sarma.prl} whereas finite temperature
Quantum Monte Carlo techniques propose it to be a
metal\cite{scalettar.prl}.

Another theoretical model for the study of the band to Mott insulator
crossover is the bilayer Hubbard model (BHM). The presence of
copper-oxide bilayers, well described by doped Hubbard planes, in many
of the high-$T_c$ materials adds to the practical importance of
studying the bilayer Mott system theoretically. The BHM was studied
recently using Dynamical mean field theory (DMFT),\cite{fuhrmann} and
a smooth crossover between band and Mott insulators was observed.
Since this study dealt with the infinite dimensional lattice,
intralayer spatial correlations are absent.  As a two-dimensional
Hubbard model exhibits fundamentally different behavior from the
infinite-dimensional Hubbard model, such as a pseudogap\cite{kyung}
and $d$-wave superconductivity\cite{sarma} upon doping, a study
incorporating the two-dimensional nature of the Hubbard planes in the
BHM is expected to show richer physics.

In this work, we study the evolution of an antiferromagnetic (AF) Mott
insulator into a band insulator as the interlayer hopping is varied in
a bilayer Hubbard model at half-filling using cluster dynamical mean
field theory\cite{cdmft}. With interplane hopping set to $t_\perp$,
the Hamiltonian for the bilayer Hubbard model can be written as \beqa
H \!&=&\! -t\sum_{\langle i j\rangle\sigma \alpha } \Bigl(
c_{i\sigma\alpha}^{\dagger}c_{j\sigma\alpha} + H.c.\Bigr)
+U\sum_{i\alpha}n_{i\downarrow\alpha}n_{i\uparrow\alpha} \nonumber \\
&&\!
-t_\perp\sum_{i\sigma\alpha}c_{i\sigma\alpha}^{\dagger}c_{i\sigma
(1-\alpha)} .\eeqa
Here $\alpha=0,1$ labels the two Hubbard planes and $U$ represents the
onsite Coulomb repulsion. $c_{i\sigma\alpha}$ denotes the destruction
operator at site $i$, with spin $\sigma$ and on plane
$\alpha$. Chemical potential is chosen as $\mu=U/2$ so that the
density is fixed to unity. In the limit $U\rightarrow0$, it can be
seen that the system goes from an uncorrelated metal to a band
insulator at $t_\perp=4t$ when the splitting between the bonding and
antibonding bands produced by the two layers results in a finite
gap. At large $U$ and $t_\perp\rightarrow0$, the system decouples into
two Hubbard systems with a finite Mott gap and AF long range order in
each layer.

In the recent study on the infinite-dimensional BHM, the two-plane
system was reduced to two impurities embedded in a self-consistently
determined bath\cite{fuhrmann}. For $t_\perp=0$ the authors found a
phase transition at $U=U_c$ from a metallic phase to a Mott insulating
phase. As $t_\perp$ is switched on, the Mott insulator (MI) evolves
continuously into a band insulator (BI) with no sign of a phase
transition. In what follows, we shall argue that in-plane spatial
correlations provide a competing energy scale that is missing in the
infinite dimension. Taking both intraplane and interplane spatial
correlations into account within CDMFT we find that for $U<8t$ as
$t_\perp$ is increased from zero the MI goes through a metallic phase
before entering a band insulating phase. For larger values of $U$, the
system undergoes a direct crossover from a MI to a BI without an
intervening metallic phase.

\begin{figure}[tb]
    \includegraphics[width=6.0cm]{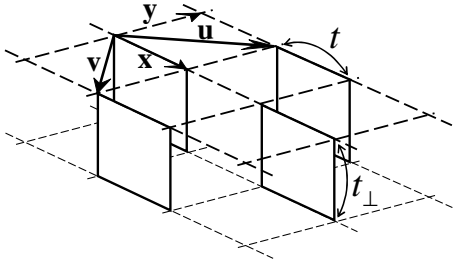}
  \caption{Tiling the infinite bilayer with a $2 \times 2$ cluster}
\label{tiling}
\end{figure}

CDMFT is a non-perturbative technique where the full many-body problem
is mapped onto local degrees of freedom treated exactly within a
finite cluster that is embedded in a self-consistent
bath\cite{cdmft,maier}. It is a natural generalization of single-site
DMFT\cite{georges} to incorporate spatial correlations. The method has
passed rigorous tests for the 1D Hubbard model where it compares well
to exact solutions\cite{cdmftapp} and has been applied to a variety of
problems\cite{kyung,sarma}. The first step in the CDMFT method
involves a tiling of the infinite lattice by a finite cluster.  As
seen in Fig.~\ref{tiling}, we consider a $2 \times 2$ tiling cluster
which includes both the in-plane and interplane hopping processes with
a unit cell defined by the vectors $\bf{u=x+y}$ and $\bf{v=x-y}$.
Using CDMFT, the BHM on the infinite lattice reduces to the
cluster-bath Hamiltonian below, that is subject to a self-consistency
condition: \beqa H_c \!&=&\!\! \sum_{\langle\mu\nu\rangle\sigma}
t_{\mu\nu} \bigl( c_{\mu\sigma}^{\dagger}c_{\nu\sigma} + H.c. \bigr) +
U\sum_\mu n_{\mu\uparrow}n_{\mu\downarrow} \nonumber \\ &+&\!\!
\sum_{m\sigma}\epsilon_{m\sigma}
a_{m\sigma}^{\dagger}a_{m\sigma}+\sum_{m\mu\sigma}V_{m\mu\sigma}(a_{m\sigma}^{\dagger}c_{\mu\sigma}+H.c).
\label{clusterham}
\eeqa 
Here, $\mu,\nu=1,\cdot\cdot\cdot, N_c$ denote indices labeling the
cluster sites and $m=1,\cdot\cdot \cdot,N_b$ represent those in the
bath. The self-consistent calculation proceeds by an initial guess for
the cluster-bath hybridization $V_{m\mu\sigma}$ and the bath site
energies $\epsilon_{m\sigma}$ to obtain the cluster Green's function
$G_c^{\mu\nu}$. Applying the Dyson's equation,
${\bf \Sigma}_c={\bf G}_{0}^{-1}-{\bf G}_{c}^{-1}$, where ${\bf G}_{0}$ denotes the
non-interacting Green's function for the Hamiltonian in
Eq.~(\ref{clusterham}), the cluster self-energy is obtained.  ${\bf \Sigma}_c$
is then used in the self-consistency condition below to determine the
local Green's function for the lattice: \beq
{\bf G}_{loc}(z)=\frac{N_c}{2\pi^2} \int d{\bf P_u}
d{\bf P_v}\frac{1}{z+\mu-t({\bf P_u},{\bf P_v})-{\bf \Sigma}_c(z)}
\label{selfcon}
\eeq Here, $\bf{P_u}$ and $\bf{P_v}$ denote reciprocal lattice vectors
conjugate to the spatial unit vectors $\bf{u}$ and $\bf{v}$
respectively, while $z=i\omega_n$ is the Fermionic Mastubara
frequency. ${\bf G}_{loc}$ can be used to obtain a new Weiss field,
${\bf G}_{0}$, and consequently, a new set of $V_{m\mu\sigma}$ and
$\epsilon_{m\sigma}$'s by means of a conjugate gradient minimization
program to close the iterative loop. In this work, the Lanczos method
is used to solve the cluster-bath Hamiltonian and we fix the bath size
to $N_b=8$. The Lanczos method can access both the strong and weak
coupling regimes with equal ease and it is well suited to compute
dynamical quantities directly in real frequency. Rotational symmetries
of the cluster on a square lattice, together with particle-hole
symmetry at half-filling, reduce the number of bath parameters
significantly. These bath parameters can depend on spin, allowing for
symmetry breaking solutions.  For details of the method, we refer to
earlier work\cite{cdmft,cdmftapp}.

\begin{figure}[tb]
  \begin{center}
    \includegraphics[width=8.5cm]{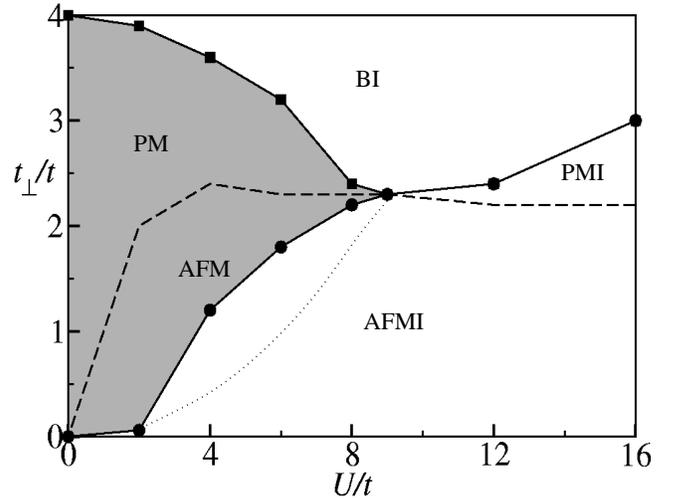}
  \end{center}
  \caption{Phase diagram for bilayer Hubbard model. Shaded portion
  shows the extent of the metallic region (paramagnetic metal (PM) and
  antiferromagnetic metal (AFM)) in the presence of magnetic
  order. Dashed line showing the magnetic phase boundary extends into
  the insulating region separating the paramagnetic Mott insulator
  (PMI) from the antiferromagnetic Mott insulator (AFMI). Suppression
  of magnetic order extends the metallic phase down to lower $t_\perp$
  shown by the dotted curve.}
\label{phasediag}
\end{figure}

Let us start from the phase diagram with suppression of the magnetic
order.  Numerical results are shown in Fig.~\ref{phasediag}.  We
observe three states; a Mott insulating state at large $U$, a
band insulating state at large $t_\perp$, and a metallic state in
between. At $U=0$, the BI and the metallic states are separated at
$t_\perp=4$ where the bonding and antibonding bands are separated.
Critical $t_\perp$ decreases with increasing $U$ because the width of
bonding and antibonding bands broaden due to correlation.  Above
$U=12$, the BI and the MI states are adiabatically connected, but by
close inspection of the charge gap, two regions are distinguishable;
in the MI (BI) phase, $\Delta_c$ decreases (increases) with increasing
$t_\perp$. Those features are consistent with the DMFT result on the
infinite-dimensional BHM.  In contrast to the infinite-dimensional
case, the MI state extends down to small $U$ and $t_\perp$ in the
two-dimensional case because of the intralayer spatial correlation,
and the system passes through two phase transitions when increasing
$t_\perp$ at $0<U<12$.  Thus, the two-dimensional BHM provides
topologically distinct behavior as compared to the
infinite-dimensional one. As will be discussed in detail below, this
essential feature is not altered even by including the magnetic
ordering.
\begin{figure}[tb]
  \begin{center}
    \includegraphics[width=7.5cm]{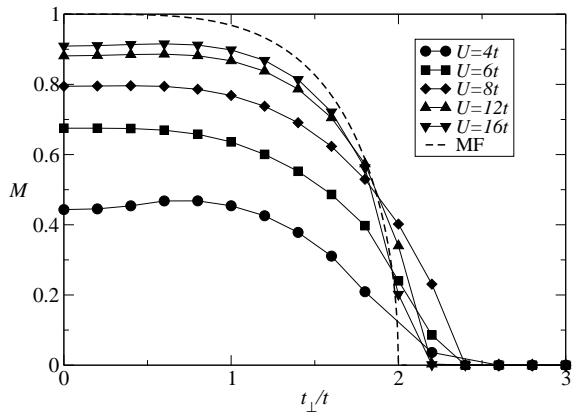}
  \end{center}
  \caption{Staggered magnetization a function of $t_\perp$ for $U=4t,6t,8t,12t,16t$} 
\label{mag}
\end{figure}

We next investigate the magnetic properties of the bilayer Hubbard
model in detail.  The staggered magnetization
$M=n_\uparrow-n_\downarrow$ is plotted as the interplane hopping is
varied in Fig.~\ref{mag} for different values of $U$. At $t_\perp=0$,
the two planes are decoupled and the value of the staggered
magnetization increases as $U$ is increased, but quickly saturates to
a value below unity because of quantum fluctuations. With increasing
$t_\perp$, $M$ stays fairly constant until $t_\perp=t$ and then drops
rapidly to zero at about $t_\perp=2t$. This value of $t_\perp=2t$ is
universal and does not depend on $U$ for all $U>8t$. The vanishing of
the staggered magnetization at small $U$ is caused by an increase in
the effective bandwidth and a consequent reduction in the exchange
coupling.  In contrast, critical value $t_\perp \approx 2t$ at large
$U$ is due to the instability toward the formation of a local singlet.
In this limit, the BHM is reduced to the $S=1/2$ Heisenberg model with
the intralayer exchange $J=4t^2/U$ and the interlayer exchange
$J_\perp=4t_\perp^2/U$.  By taking a dimer on the interlayer bond as a
unit and introducing a mean-field decoupling on the intralayer bonds,
critical exchange coupling for the transition from a N{\'e}el ordered
state to an interlayer dimerized state is obtained as $J_\perp=4J$ at
$T=0$, thus, $t_\perp=2t$.  Staggered magnetization $M=2\langle S_z
\rangle$ at $t_\perp<2t$ is also computed. As shown with a dashed line
in Fig.~\ref{mag}, this mean-field approximation to the bilayer
Heisenberg model gives reasonable agreement with CDMFT at large $U$.
\begin{figure}[tb]
  \begin{center}
    \includegraphics[width=6.5cm]{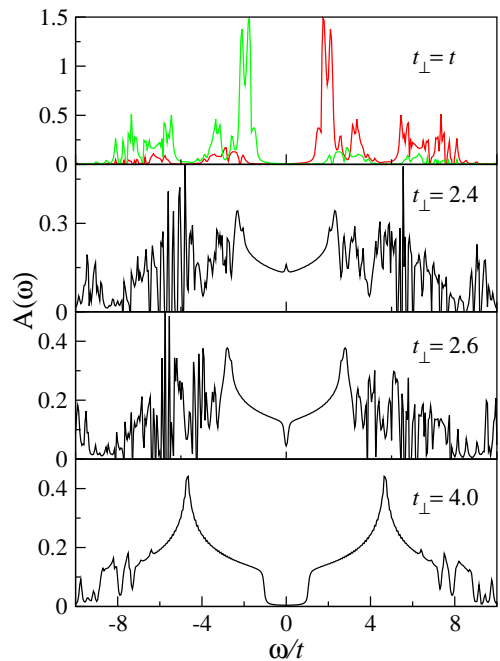}
  \end{center}
  \caption{Local density of states as $t_\perp$ is varied at $U=8t$.
  Spin-resolved density of states shown for $t_\perp=1.0$ where
  magnetic order exists. Note the metallic state at $t_\perp=2.4t$.}
\label{ldos}
\end{figure}

Next, we look at the local density of states for the BHM as a function
of $t_\perp$ at $U=8t$ as shown in Fig.~\ref{ldos}. As $t_\perp$ is
increased the Mott gap reduces till it closes completely at
$t_\perp=2t$ to give rise to a metallic state. The LDOS shows a
clear quasiparticle peak along with the lower and upper Hubbard bands
characteristic of Mott physics at $t_\perp=2.4t$. Upon further
increase of $t_\perp$ we see that a gap opens abruptly to reveal a
band insulating state. Increasing $t_\perp$ in the band insulating
phase causes the charge gap to grow monotonously unlike the Mott
insulating phase where it reduces to zero.
\begin{figure}[tb]
  \begin{center}
    \includegraphics[width=7.5cm]{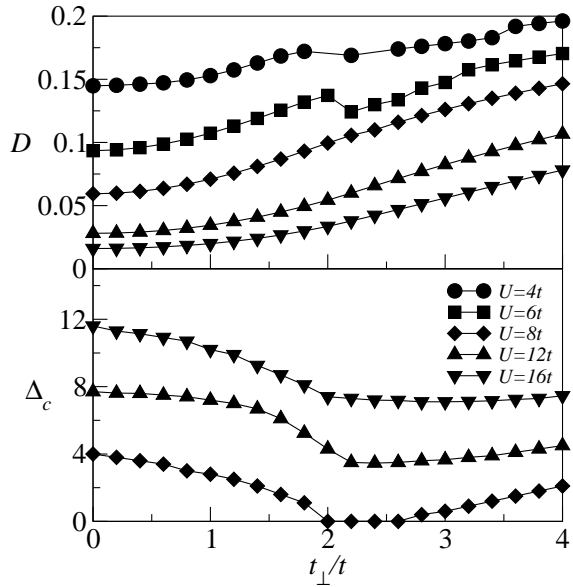}
  \end{center}
  \caption{Double occupancy (top panel) and charge excitation gap (bottom) as 
  functions of $t_\perp$ for various $U$ indicated. } 
\label{double-gap}
\end{figure}
In Fig.~\ref{double-gap} we plot the double occupancy and the charge
gap as a function of $t_\perp$ for different values of $U$. For
$U<8t$, as $t_\perp$ is increased in the Mott phase, the double
occupancy rises and then shows a sharp kink as the system enters a
metallic phase and also upon going into a band insulating phase. For
larger values of $U$, as expected, the crossover from a Mott insulator
to a band insulator is smooth with the double occupancy showing no
signs of a kink. The charge gap for large $U$ shows a continuous
decrease in the Mott phase but never goes to zero. The crossover to a
band insulator around $t_\perp=2t$ is easily identified by the
increase in the charge gap with $t_\perp$.

The phase diagram for the bilayer Hubbard model that emerges for our 
calculation is shown in Fig.~\ref{phasediag}. 
We identify a region in the phase 
diagram which shows an intermediate metallic region, depicted by the
shaded portion, which separates the band and Mott insulating
phases. The width of the metallic region narrows with increasing $U$
and eventually for $U>8t$ there is a direct crossover from a band
insulator to a Mott insulator. The magnetic phase boundaries are
depicted by the dashed line which separates a antiferromagnetic metal
from a paramagnetic metal. In the insulating portion of the phase
diagram the magnetic phase boundary separates the antiferromagnetic
Mott insulator (AFMI) from a paramagnetic Mott insulator (PMI). 
As we discussed earlier, emergence of intermediate metallic region separating BI and MI 
is a generic behavior in two-dimensional BHM independent of magnetic ordering. 

In summary, we have obtained the zero temperature phase diagram of the
bilayer Hubbard model at half-filling using cluster dynamical mean
field theory. 
We computed the double occupancy and the local density of states as a function of the
interlayer coupling $t_\perp$ and the Coulomb repulsion $U$ to
identify a clear demarcation between a Mott insulating phase and a 
band insulating phase as the interlayer coupling is varied. 
For weak to intermediate values of interaction ($U<12t$), the Mott insulator at 
small $t_\perp$ is separated from the band insulator at large
$t_\perp$ by an intermediate metallic phase. For $U>12t$, there is a
direct crossover from a Mott insulator to a band insulator as
$t_\perp$ is increased. This crossover is most clearly manifest in the
behavior of the charge gap which first reduces in the Mott phase with
increasing $t_\perp$ and then increases in the band insulating
phase. 
These results should be contrasted with the DMFT results for the
infinite-dimensional model which always find a smooth crossover
between the BI and MI states.\cite{fuhrmann} These differences arise
due to the presence of in-plane spatial correlations in the
two-dimensional model stabilizing a Mott phase for all $U$ at
half-filling which is accessible by cluster DMFT.
We completed the phase diagram by further allowing for magnetic order.
The phase diagram is essentially unchanged, except that the onset of
the metallic phase occurs at a larger value of $t_\perp$ for a fixed
value of $U$.

S.S.K acknowledges discussions with Achim Rosch. S.S.K. is supported
by the LDRD program at Oak Ridge National Laboratory. This work is
supported by the Division of Materials Sciences and Engineering, Office
of Basic Energy Sciences, U.S. Department of Energy, under contract
DE-AC05-00OR22725 with Oak Ridge National Laboratory, managed and
operated by UT-Battelle, LLC.

\end{document}